\documentclass{article} 

\usepackage{amsmath,amssymb,amsfonts}
\usepackage{ifpdf}
\ifpdf
  \usepackage{graphicx}
  \usepackage[hidelinks]{hyperref}
\else
  \usepackage[dvipdfmx]{graphicx}
  \usepackage[dvipdfmx,hidelinks]{hyperref}
\fi
\usepackage{booktabs}
\usepackage{tikz}
\usepackage{flafter}

\usepackage{geometry}
\usepackage{caption}
\usepackage{subcaption}
\usepackage{float}

\graphicspath{{figures/}}

\title{Differential Machine Learning for 0DTE Options with Stochastic Volatility and Jumps}

\author{Takayuki Sakuma\footnote{e-mail: tsakuma@soka.ac.jp. }\\
Faculty of Economics and Business Administration, Soka University}
\date{\today}

\begin{document}
\maketitle

\begin{abstract}
We present a differential machine learning method for zero-days-to-expiry (0DTE) options under a stochastic-volatility jump-diffusion model. To handle the ultra-short-maturity regime, we express the option price in Black--Scholes form with a maturity-gated variance correction, combining supervision on prices and Greeks with a PIDE-residual penalty. Prices and Greeks are derived from a single trained pricing network, while jump-term identifiability is ensured by a jump-operator network fitted jointly in a three-stage procedure. The method improves jump-term approximation relative to one-stage baselines while maintaining comparable pricing errors. Furthermore, it reduces errors in Greeks, produces stable one-day delta hedges, and offers significant speedups over Fourier-based benchmarks. Calibration experiments demonstrate the network's efficiency as a pricer and incorporating jump-intensity price sensitivity into the learning process further improves the overall model fit. We also consider a jump rough Heston model.
\end{abstract}

\section{Introduction}
Zero-days-to-expiry (0DTE) options have grown rapidly over the past few years and now account for a large share of trading volume.
Recent empirical work documents frequent intraday jumps in 0DTE markets~\cite{Bozovic2025}, studies the role of gamma-related open-interest exposures~\cite{DimErakerVilkov2024}, and analyzes the pricing implications of short-horizon tail risk in 0DTE options~\cite{BandiFusariReno2023}. These findings raise two practical challenges: first, the underlying dynamics may be better described by a diffusion model with jumps; second, the very short maturities and frequent intraday rebalancing demand fast computation of option prices and Greeks.

We apply differential machine learning (DML)~\cite{hugeSavine2020dml,frandsen2022deltaforce} to 0DTE options under the Bates stochastic-volatility jump-diffusion (SVJD) model. Learning-based models can reduce pricing costs: once trained, prices and Greeks are obtained from a single network evaluation. However, the numerically most challenging region is near the money, where Greeks can become very large for 0DTE options with jumps. A common approach in machine-learning PDE solvers is to enforce the governing PDE by penalizing its residual at sampled state points, together with terminal and boundary conditions~\cite{raissi2019pinn}. Several recent studies apply this residual-penalty approach to option pricing with jumps by including a PIDE-residual term in the training objective~\cite{fu2020unsupervised,sun2025pinnmerton,bansal2026pinnpide}. Three design choices are central to our approach:
\begin{enumerate}
  \item We adopt DML, training a single price network jointly on option values and Greeks. The Greeks are obtained by automatic differentiation of the network output with respect to its inputs and enter the loss directly.
  \item Instead of predicting prices directly, the network learns a variance correction inside a Black--Scholes formula \cite{black1973pricing}, scaled so that the correction vanishes as $\tau\to 0$. This preserves the correct short-maturity limit (the payoff) and reduces the approximation burden in the near-singular region. More broadly, this design is related to neural-network surrogates for option pricing and implied-volatility computation \cite{liu2019pricing}, and to implied-volatility smoothing methods that impose no-arbitrage structure to improve surface regularity \cite{ackerer2020deep}.
  \item We introduce a second neural network to represent the compensated jump operator. If the jump component is identified only through a PIDE-residual penalty, the optimizer can trade off diffusion and jump errors while keeping the overall residual small. A small residual therefore does not guarantee that the learned jump operator matches the model-implied jump integral. The second network makes the residual penalty informative about the jump contribution.
\end{enumerate}
Recent calibration studies using DML fall into two broad strands. Sridi and Bilokon (2023) apply DML to the calibration of vanilla European puts under the Heston model, while Polala and Hientzsch (2023) extend parametric DML to joint pricing and calibration problems~\cite{SridiBilokon2023,PolalaHientzsch2023}. In these approaches, DML primarily replaces the pricer to gain speed. Zhang et al. (2025), by contrast, propose a gradient-based scheme that learns both prices and sensitivities with respect to model parameters~\cite{ZhangAmiciMorandotti2025}. We ask whether our approach can play both roles: as a fast pricer in calibration and as a way to improve calibration in a focused extension. We also ask whether our DML framework can be extended to rough volatility models. To this end, we consider a jump rough Heston model based on a multi-factor Markovian approximation, which replaces the fractional kernel by a finite sum of exponentials~\cite{AbiJaberElEuch2019}.

\paragraph{Paper organization.} Section~\ref{sec:bates} presents the Bates stochastic-volatility jump-diffusion model. Section~\ref{sec:dml-jump} introduces the DML-based neural network and the three-stage training scheme. Section~\ref{sec:nn-training} specifies the loss functions and constraints. Section~\ref{sec:numerical} reports numerical experiments, including calibration exercises. 

\section{Bates model}\label{sec:bates}
We work with the Bates stochastic-volatility jump-diffusion (SVJD) model, which combines a Heston-type variance process with Merton-style lognormal price jumps \cite{Merton1976,Heston1993,Bates1996}. For brevity, we refer to this simply as the Bates model throughout the paper.
Under the risk-neutral measure, let $S_t$ denote the underlying asset, $V_t$ the instantaneous variance, $r$ the risk-free rate, and $q$ the dividend.
The dynamics are
\begin{align}
  \frac{dS_t}{S_{t^-}} &= (r - q - \lambda \kappa_J)\,dt + \sqrt{V_t}\,dW^S_t + (e^{Y} - 1)\,dN_t, \label{eq:SDE-S} \\
  dV_t &= \kappa(\theta - V_t)\,dt + \sigma_v \sqrt{V_t}\,dW^V_t, \label{eq:SDE-V}
\end{align}
where $(W^S_t, W^V_t)$ is a two-dimensional Brownian motion with correlation $\rho$, $N_t$ is a Poisson process with intensity $\lambda$, and the jump sizes are i.i.d.\ with $Y \sim \mathcal{N}(\mu_J,\sigma_J^2)$.
Here $\kappa$ is the variance mean-reversion speed in \eqref{eq:SDE-V} and $\kappa_J = \mathbb{E}[e^Y - 1] = \exp(\mu_J + \tfrac{1}{2}\sigma_J^2) - 1$.

For a European call option with maturity $\tau$ and strike $K$, the risk-neutral call price $C$ satisfies the following PIDE:
\begin{align}
  \frac{\partial C}{\partial \tau}
  &= (r - q)S\frac{\partial C}{\partial S}
   + \kappa(\theta - V)\frac{\partial C}{\partial V}
   + \frac{1}{2} V S^2\frac{\partial^2 C}{\partial S^2}
   + \frac{1}{2}\sigma_v^2 V\frac{\partial^2 C}{\partial V^2}
   + \rho\sigma_v V S\frac{\partial^2 C}{\partial S\partial V}
   - rC
   + \lambda\,\mathcal{J}[C],
   \label{eq:PIDE-S}
\end{align}
with terminal condition $C(S,V,0)=(S-K)^+$.
The compensated jump operator is
\begin{equation}
  \mathcal{J}[C](S,V,\tau)
  = \int_{\mathbb{R}}
  \Big[
    C(Se^{y},V,\tau)-C(S,V,\tau)-(e^{y}-1)S\,C_S(S,V,\tau)
  \Big] f_Y(y)\,dy,
\end{equation}
where $f_Y$ denotes the density of the logarithmic jump size $Y$.
We use the dimensionless log-moneyness
\[ x := \log(S/K), \qquad \tau := T-t \]
and the diffusion part of the operator is
\begin{align}
  \mathcal{L}_{\mathrm{diff}} u
  &= u_t + (r-q)u_x + \kappa(\theta-V)u_V + \tfrac12 V(u_{xx}-u_x)
   + \rho\sigma_v V\,u_{xV} + \tfrac12\sigma_v^2 V u_{VV} - ru,
  \label{eq:Ldiff}
\end{align}
where $u_t=-u_\tau$ since $\tau$ is time-to-maturity.
We define the residual
\begin{equation}
  R(\mathbf{x}) := \mathcal{L}_{\mathrm{diff}} u_\phi(\mathbf{x}) + \lambda\,J_\psi(\mathbf{x}),
  \label{eq:residual}
\end{equation}
where $J_\psi$ denotes a neural approximation to the normalized compensated jump operator
\begin{equation}
J_\psi(\mathbf{x})
:= \int_{\mathbb{R}}
\Big[
u(x+y,V,\tau)-u(x,V,\tau)-(e^y-1)u_x(x,V,\tau)
\Big] f_Y(y)\,dy.
  \label{eq:residual_jump}
\end{equation}
In the architecture below, the second network outputs $J_\psi$ directly and is supervised against a numerical quadrature proxy for $J_\psi(\mathbf{x})$.

\section{DML for PIDEs}\label{sec:dml-jump}

\begin{figure}[t]
  \centering
  \resizebox{\linewidth}{!}{
\begin{tikzpicture}[
  >=latex,
  scale=0.95,
  every node/.style={transform shape},
  box/.style={draw, rounded corners, minimum height=8mm, align=center, font=\footnotesize},
  input/.style={box, fill=yellow!35},
  price/.style={box, fill=blue!10},
  jump/.style={box, fill=green!15},
  outbox/.style={box, fill=orange!20},
  note/.style={box, fill=gray!10, font=\scriptsize},
  stage/.style={note, text width=2.25cm, align=center, minimum height=12mm}
]

\node[input, text width=2.2cm] (x) at (-1.5,-0.5)
{Inputs\\[-0.2em]{$\mathbf{x}$}};

\node[price] (pnet) at (2.2,0.9) {Variance-correction network\\$\Delta V_\phi(\mathbf{x})$};
\node[jump]  (jnet) at (2.2,-1.9) {Jump-operator network};

\node[outbox, align=left, text width=5.45cm] (bs) at (8.05,0.9) {BS with a variance adjustment\\[-0.25em]
{\scriptsize
$g(\tau)=1-\exp(-\tau/\tau_0)\;\;$\\
$V_{\mathrm{eff}}=\max\{V+g(\tau)\,\Delta V_\phi(\mathbf{x}),\,\varepsilon\}$\\
$u_\phi(\mathbf{x})=C_{\mathrm{BS}}(S=e^{x},K=1,\tau,\sqrt{V_{\mathrm{eff}}})$}};
\node[outbox, text width=2.2cm] (jout) at (6.55,-1.9) {Compensated\\jump term\\$J_\psi(\mathbf{x})$};

\node[price, text width=2.7cm] (greeks) at (9.55,-1.9) {Greeks\\$(delta,gamma,vega)$};

\node[note, minimum width=7.6cm, text width=7.6cm] (res) at (4.4,-3.75)
{PIDE residual:\\
$R(\mathbf{x})=\mathcal{L}_{\mathrm{diff}}u_\phi(\mathbf{x})+\lambda\,J_\psi(\mathbf{x})$\;\;\\
};

\draw[->] (x) -- (pnet);
\draw[->] (x) -- (jnet);
\draw[->] (pnet) -- (bs);
\draw[->] (jnet) -- (jout);



\draw[->, dashed] (bs.south) -- ++(0,-0.25) -| (greeks.north);

\draw[->, dashed] (bs.south) -- ++(0,-0.25) -| (res.north);
\draw[->, dashed] (jout.south) -- ++(0,-0.25) -| (res.north);

\draw[->, dashed] (greeks.south) -- ++(0,-0.25) -| (res.north);

\node[stage] (s1) at (1.6,-5.5)
{\textbf{Stage 1}\\
train $\phi$\\
freeze $\psi$};
\node[stage] (s2) at (5.0,-5.5)
{\textbf{Stage 2}\\
train $\psi$ \\
freeze $\phi$};
\node[stage] (s3) at (8.4,-5.5)
{\textbf{Stage 3}\\
joint fine-tune $\phi,\psi$\\
+ PIDE residual penalty\\
+ self-consistency};
\draw[->] (s1) -- (s2);
\draw[->] (s2) -- (s3);

\end{tikzpicture}}
  \caption{Architecture and training procedure. Stages~1--3 describe the three-stage training scheme. A variance-correction network returns $\Delta V_\phi(\mathbf{x})$, multiplied by a deterministic maturity function $g(\tau)$ so that the correction vanishes as $\tau\to 0$. Prices are produced by substituting the resulting effective variance into the Black--Scholes call formula, yielding $u_\phi(\mathbf{x})$ (not the standard BS price unless $\Delta V_\phi\equiv 0$). A separate network outputs the compensated jump contribution $J_\psi(\mathbf{x})$. Greeks are obtained by automatic differentiation of $u_\phi$. The jump-PIDE residual $R(\mathbf{x})$ is computed and penalized at randomly sampled points. Stages~1--3 depict the three-stage training scheme used for jump-term identifiability.}
  \label{fig:dml-twin-network}
\end{figure}
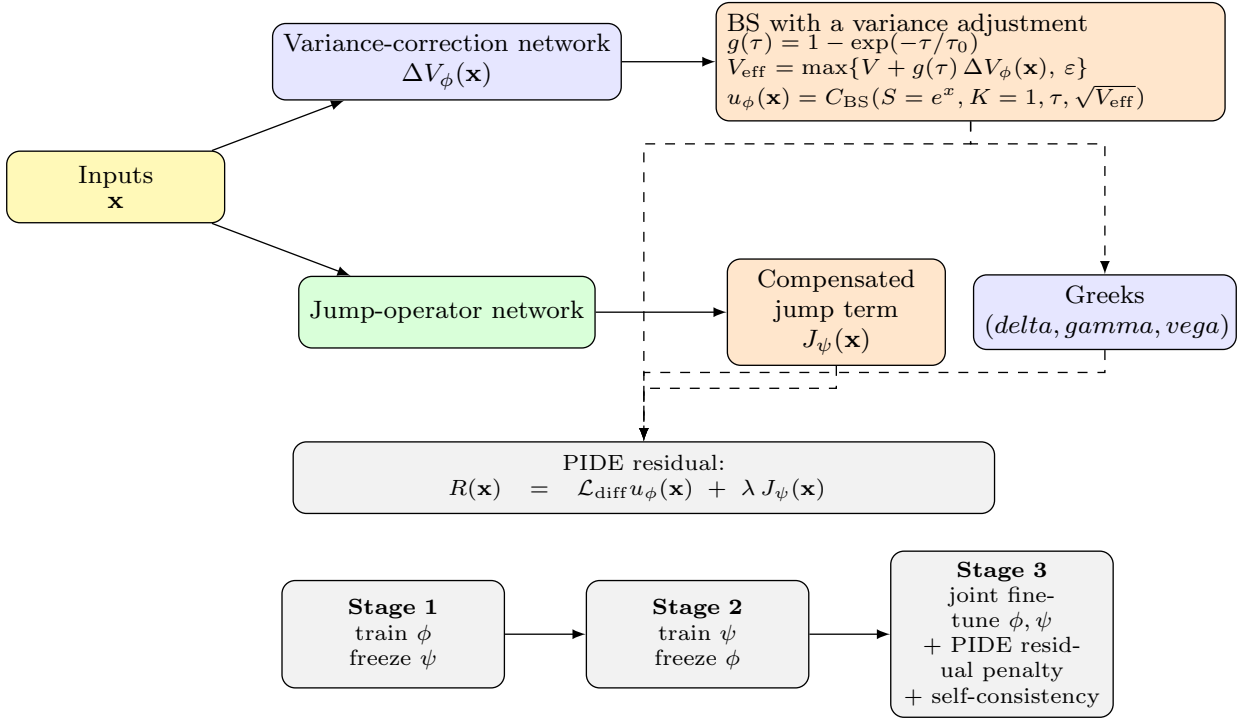

Figure~\ref{fig:dml-twin-network} summarizes the model architecture. The solid arrows correspond to forward evaluations while dashed arrows indicate computations derived from these outputs. The input is
\begin{equation}
  \mathbf{x} = (x,\tau,V,\kappa,\theta,\sigma_v,\rho,\lambda,\mu_J,\sigma_J)\in\mathbb{R}^{10}.
\end{equation}
Rather than predicting prices directly, the first network outputs a variance correction $\Delta V_\phi(\mathbf{x})$.
We define an effective variance
\begin{equation}
  V_{\mathrm{eff}}(\mathbf{x})
  = \max\big\{ V + g(\tau)\,\Delta V_\phi(\mathbf{x}),\,\varepsilon\big\},
  \qquad
  g(\tau)=1-\exp(-\tau/\tau_0),
  \label{eq:veff}
\end{equation}
and we return a Black--Scholes call price with volatility $\sigma_{\mathrm{eff}}=\sqrt{V_{\mathrm{eff}}}$. The function $g(\tau)$ forces the learned variance correction to vanish near expiry, which stabilizes training and preserves the payoff limit.
Since we work in log-moneyness, we normalize the strike to $K=1$:
\[
  u_\phi(\mathbf{x}) := C_{\mathrm{BS}}\!\big(S=e^{x},\,K=1,\,r=0.01,\,q=0,\,\tau,\,\sigma_{\mathrm{eff}}\big)
\]
and Greeks are obtained by automatic differentiation of $u_\phi$. In parallel, a jump-operator network approximates the compensated jump operator $J_\psi(\mathbf{x})$, and the two networks are coupled through the jump-PIDE residual.
Because the jump term is not identifiable from a residual loss alone, we supervise $J_\psi$ and use the three-stage schedule described in Subsection~\ref{sec:three-stage}.

\subsection{Twin network}\label{sec:dml-ident}
A standard supervised model fits a network $u_\phi(\mathbf{x})$ by
\[
\min_{\phi}\;\mathbb{E}\big[(u_\phi(\mathbf{x})-u(\mathbf{x}))^2\big].
\]
DML augments this objective by adding supervised targets for selected Greeks $\partial u/\partial x_i$:
\begin{equation}
\min_{\phi}\;\mathbb{E}\Big[(u_\phi-u)^2 + \sum_{i\in\mathcal{I}} \omega_i\big(\partial_{x_i}u_\phi - \partial_{x_i}u\big)^2\Big],
\label{eq:dml-sobolev}
\end{equation}
where $\partial_{x_i}u_\phi$ are computed by automatic differentiation and $\mathcal{I}$ indexes the Greeks of interest (e.g., delta, gamma, and vega). Following Huge and Savine (2020), we use the term ``twin network'' to denote a single price network paired with Greeks obtained by automatic differentiation.
Because the same parameters $\phi$ must explain both values and derivatives, derivative supervision supplies shape information and improves accuracy.

Training can be further regularized by adding a PDE-residual penalty computed from the same automatic-differentiation derivatives. However, if we approximate the nonlocal jump operator $J[u]$ by a separate network $J_\psi(\mathbf{x})$ and use it only through the residual
\begin{equation}
  R(\mathbf{x}) \;=\; \mathcal{L}_{\mathrm{diff}}u_\phi(\mathbf{x}) + \lambda\,J_\psi(\mathbf{x}),
  \label{eq:residual-ident}
\end{equation}
$u_\phi$ and $J_\psi$ are not separately identifiable: in principle, for any $u_\phi$ one can set
$J_\psi=-\mathcal{L}_{\mathrm{diff}}u_\phi/\lambda$ and obtain $R\equiv 0$.
This mechanism can lead to a degenerate solution in which the jump network cancels diffusion-operator errors and reduces the residual without learning a meaningful jump contribution. We therefore supervise the jump operator explicitly.
Using reference prices $u^{\text{ref}}$, we construct a numerical proxy for the jump term and train $J_\psi$ to match it.

\subsection{Three-stage training with jump supervision}\label{sec:three-stage}
As noted in Section~\ref{sec:dml-ident}, $J_\psi$ can act as a residual-cancelling degree of freedom, absorbing diffusion-operator errors without approximating the jump operator.
We therefore use a three-stage schedule:
\begin{enumerate}
  \item \textbf{Stage 1 (price and Greeks).} Train the price network while freezing $J_\psi$, using the price and Greek terms plus the no-arbitrage penalties.
  \item \textbf{Stage 2 (jump reference).} Freeze the price network and train the jump network to match a numerical proxy of the compensated jump term computed from the reference prices:
\begin{equation}
J^{\text{ref}}(x,V,\tau)
:= \mathbb{E}\big[u^{\text{ref}}(x+Y,V,\tau)-u^{\text{ref}}(x,V,\tau)\big] - \kappa_J\,u^{\text{ref}}_x(x,V,\tau).
  \label{eq:jump-comp}
\end{equation}
On randomly sampled state points, we approximate the expectation in~\eqref{eq:jump-comp} with a Gauss--Hermite rule applied to $u^{\text{ref}}(x+Y,V,\tau)$ and penalize $J_\psi-J^{\text{ref}}(x,V,\tau)$ using a Huber loss function.
  \item \textbf{Stage 3 (joint refinement).} Train both networks jointly. In this stage we retain weak supervision against $J^{\text{ref}}(x,V,\tau)$ and add a \emph{self-consistency} regularizer that penalizes the mismatch between $J_\psi$ and a low-order numerical jump evaluation based on the current price network.
\end{enumerate}

\section{Neural network and training}\label{sec:nn-training}
We use two fully-connected feedforward networks, each with width 192, depth 4, and SiLU activation functions (152{,}834 parameters in total).

\subsection{Loss function, constraints, and weights}

Let $\widehat{u}=u_\phi(\mathbf{x})$ and $(\widehat{\Delta},\widehat{\Gamma},\widehat{\nu})$ denote the corresponding Greeks.
We minimize the weighted objective
\begin{align}
\mathcal{L}(\phi,\psi)
&= \mathbb{E}\Big[ w(\mathbf{x})(\widehat{u}-u)^2 \Big]
+ \omega_G\,\mathbb{E}\Big[w(\mathbf{x})\,m(\mathbf{x})\sum_{g\in\{\Delta,\Gamma,\nu\}} \omega_g\,(\lambda_g (\widehat{g}-g))^2\Big] \nonumber\\
&\quad + \omega_R\,\mathbb{E}\Big[w(\mathbf{x})\,R(\mathbf{x})^2\Big]
+ \omega_{\text{col}}\,\mathbb{E}\Big[(\widehat{u}-u^{\text{ref}})^2\Big]
+ \omega_{\text{NA}}\,\mathbb{E}\big[\mathcal{P}_{\text{NA}}(\mathbf{x})\big],
\label{eq:loss}
\end{align}
with:
\begin{itemize}
  \item $w(\mathbf{x})$ upweights the ATM region and short maturities,
  \[
  w(\mathbf{x}) = 1 + W_{\text{ATM}}\,\mathbb{I}(|x|<0.1) + W_{\text{SHORT}}\,\mathbb{I}(\tau<0.02)
  + W_{\text{ATM\&SHORT}}\,\mathbb{I}(|x|<0.1,\tau<0.02).
  \]
  \item $m(\mathbf{x})=\mathbb{I}(u^{\text{ref}}(\mathbf{x})>10^{-4})$ drops Greek-loss contributions in deep OTM regions.
  \item $\mathcal{P}_{\text{NA}}$ encodes simple static no-arbitrage constraints:
  delta bounds ($0\le\Delta\le 1$), convexity ($u_{xx}-u_x\ge 0$), and vega monotonicity ($\nu\ge 0$), implemented as squared positive-part penalties $(\cdot)_+^2$, where $(z)_+=\max\{z,0\}$.
  \item Greek scaling uses $\lambda_g \approx 1/\sqrt{\mathbb{E}[g^2]}$ estimated once from the training set.
\end{itemize}

\paragraph{Self-consistency penalty.} During joint training we augment~\eqref{eq:loss} with a penalty that encourages the learned jump network to agree with a low-order numerical evaluation of the compensated operator applied to the current price network:
\begin{align}
\mathcal{L}_{\mathrm{SC}}
&:= \omega_{\mathrm{SC}}\,\mathbb{E}_{\text{res}}\Big[\big\lVert J_\psi(\mathbf{x})-\widehat{J}[u_\phi](\mathbf{x})\big\rVert^2\Big],
\label{eq:selfcons-loss}\\
\widehat{J}[u_\phi](\mathbf{x})
&:= \sum_{i=1}^{n_{\mathrm{SC}}} w_i\Big(u_\phi(x+\mu_J+\sigma_J z_i,V,\tau)-u_\phi(x,V,\tau)\Big)
- \kappa_J\,u_{\phi,x}(x,V,\tau),
\label{eq:selfcons-proxy}
\end{align}
where $(z_i,w_i)_{i=1}^{n_{\mathrm{SC}}}$ are nodes and weights of a Gauss--Hermite rule for $Z\sim\mathcal{N}(0,1)$, and we use $\omega_{\mathrm{SC}}=0.05, n_{\mathrm{SC}}=16$.

\paragraph{Robust $\Gamma$ (gamma) loss and numerical filtering.} For the most sensitive derivative targets we may replace the squared error in~\eqref{eq:loss} with a robust Huber loss and ignore samples where the reference violates convexity due to numerical quadrature error. We adopt a Huber loss to reduce the impact of isolated quadrature-induced outliers, which is especially relevant for $\Gamma$ and the jump-term. For $\Gamma$ this takes the form
\begin{align}
\mathcal{L}_{\Gamma}
&:= \mathbb{E}_{\text{data}}\Big[w(\mathbf{x})\,m(\mathbf{x})\,\mathbb{I}(\Gamma^{\text{ref}}(\mathbf{x})\ge 0)\,\mathrm{Huber}_\delta\big(\lambda_\Gamma(\widehat{\Gamma}(\mathbf{x})-\Gamma^{\text{ref}}(\mathbf{x}))\big)\Big].
\label{eq:gamma-robust}
\end{align}
The training stages correspond to different restrictions of the full objective:
\begin{itemize}
  \item \textbf{Stage 1:} minimize~\eqref{eq:loss} using only the price and Greek terms and $\mathcal{P}_{\text{NA}}$ (i.e.\ set $\omega_R=\omega_{\text{col}}=0$ and freeze $\psi$).
  \item \textbf{Stage 2:} freeze $\phi$ and minimize a Huber loss on the jump network, $\mathbb{E}[\mathrm{Huber}(J_\psi-J^{\text{ref}}(x,V,\tau))]$, where $J^{\text{ref}}$ is computed from the reference Fourier pricer (Section~\ref{sec:three-stage}).
  \item \textbf{Stage 3:} jointly refine both networks with the full loss~\eqref{eq:loss} (small $\omega_R$) plus weak jump supervision and the self-consistency penalty~\eqref{eq:selfcons-loss}. For the $\Gamma$ target we use the robust variant~\eqref{eq:gamma-robust}.
\end{itemize}

\section{Numerical experiments}\label{sec:numerical}
Our experiments address four questions: how DML and residual regularization affect 0DTE price/Greek accuracy; whether the learned jump term is actually identified; whether the resulting prices and Greeks remain reliable in hedging tests; and whether a parameter-gradient extension can improve calibration. Additional model-choice robustness checks (BS/Merton baselines and an SVCJ extension) are reported in Appendix~\ref{app:modelchoice}.

\subsection{Setting}\label{sec:numerical-data-generation}
Benchmark prices and Greeks are generated by a Fourier-transform pricer for the Bates model (1024-point quadrature; cutoff $u_{\max}=1000$) \cite{CarrMadan1999}.
Reference Greeks are computed by automatic differentiation of the same implementation.
\begin{itemize}
  \item Log-moneyness: with probability $p_{\text{core}}=0.7$ we draw from a near-ATM band $x\sim\mathrm{Unif}[-0.1,0.1]$ (``ATM core''); otherwise $x\sim\mathrm{Unif}[-0.5,0.5]$.
  \item Maturity (strict 0DTE): we oversample the shortest maturities. With probability $p_{\text{short}}=0.7$ we draw $\tau\sim\mathrm{Unif}[10^{-4},1/504]$; otherwise $\tau\sim\mathrm{Unif}[1/504,1/252]$.
  Here $p_{\text{short}}$ is the mixture weight for that oversampling step.
  \item Parameters are sampled uniformly over the following ranges:
  \begin{align*}
    v_0\in[0.01,0.2],\;\kappa\in[1,5],\;\theta\in[0.02,0.1],\\
    \sigma_v\in[0.1,1.0],\;\rho\in[-0.9,-0.3],\;\lambda\in[0.1,2.0],\\
    \mu_J\in[-0.2,0.0],\;\sigma_J\in[0.05,0.5].
  \end{align*}
\end{itemize}
To stabilize jump-term training, we augment the sampled states by pushing points through the jump map $x\mapsto x+Y$ and clamping to $|x|\le 6$.
We compute reference prices on these jump-shifted (``margin'') points and add a small auxiliary price-consistency loss. This extends supervision beyond $|x|\le 0.5$ to the region visited by the jump integral.
Deep OTM options have tiny prices and noisy Greeks, so we ignore Greek targets when the reference price is below $10^{-4}$.

We use $\omega_G=0.7$, $\omega_R=0.01$, $\omega_{\text{col}}=0.1$, $\omega_{\text{NA}}=0.05$, and $(\omega_{\Delta},\omega_{\Gamma},\omega_{\nu})=(1.0,0.3,0.5)$. We also set the self-consistency weight $\omega_{\mathrm{SC}}=0.05$ with $n_{\mathrm{SC}}=16$. Finally, we set $(W_{\text{ATM}},W_{\text{SHORT}},W_{\text{ATM\&SHORT}})=(1.0,1.0,1.0)$.

In the ultra-short regime, second derivatives are numerically unstable. We therefore (i) ignore deep-OTM Greek targets, (ii) use robust losses and/or clipping for sensitive targets (\(\Gamma\) and the jump term), and (iii) add no-arbitrage shape penalties, particularly convexity \((u_{xx}-u_x\ge 0)\). 

\subsection{Accuracy across models}
We compare four models:
\begin{enumerate}
  \item \textbf{A (price-only)}: train on prices only (no Greek loss, no residual penalty).
  \item \textbf{B (DML)}: train on prices and Greeks (via automatic differentiation), with no residual penalty.
  \item \textbf{C (DML+residual)}: train on prices, Greeks, and the residual penalty~\eqref{eq:loss}.
  \item \textbf{D (three-stage)}: three-stage training with jump supervision (Section~\ref{sec:three-stage}).
\end{enumerate}
We train the one-stage variants (Models~A--C) for 100 epochs with Adam (learning rate $10^{-3}$) and batch size 64. The three-stage model uses 100 epochs (Stage~1), 60 epochs (Stage~2), and 30 epochs (Stage~3).
Table~\ref{tab:variant-global} reports global RMSE, and Table~\ref{tab:variant-atm} reports RMSE in the region $|x|<0.05$ and $\tau\le 1/252$. Residual statistics are in Table~\ref{tab:variant-resid}.

\begin{table}[t]
  \centering
  \caption{RMSE (price and Greeks) on the validation set ($N=500$, seed=42).}
  \label{tab:variant-global}
  \begin{tabular}{lcccc}
    \toprule
    Model & Price RMSE & $\Delta$ RMSE & $\Gamma$ RMSE & Vega RMSE \\
    \midrule
    A (price-only) & 0.000190 & 0.00853 & 3.28 & 0.001155 \\
    B (DML) & 0.000234 & 0.00553 & 3.11 & 0.000328 \\
    C (DML+residual) & 0.000234 & 0.00554 & 3.11 & 0.000334 \\
    D (three-stage) & 0.000231 & 0.00548 & 3.11 & 0.000321 \\
    \bottomrule
  \end{tabular}
\end{table}

\begin{table}[t]
  \centering
  \caption{RMSE in the strict short-dated ATM bucket ($|x|<0.05$, $\tau\le 1/252$), computed on the validation sample ($N=500$, seed=42).}
  \label{tab:variant-atm}
  \begin{tabular}{lccc}
    \toprule
    Model & Price RMSE & $\Delta$ RMSE & $\Gamma$ RMSE \\
    \midrule
    A (price-only) & 0.000185 & 0.01272 & 4.14 \\
    B (DML) & 0.000280 & 0.00770 & 3.82 \\
    C (DML+residual) & 0.000280 & 0.00770 & 3.82 \\
    D (three-stage) & 0.000274 & 0.00761 & 3.81 \\
    \bottomrule
  \end{tabular}
\end{table}

\begin{table}[t]
  \centering
  \caption{Residual statistics of the PIDE check: mean absolute residual $\mathbb{E}|R|$ and $\mathrm{sd}(R)$.}
  \label{tab:variant-resid}
  \begin{tabular}{lcc}
    \toprule
    Model & $\mathbb{E}|R|$ & $\mathrm{sd}(R)$ \\
    \midrule
    A (price-only) & 0.0339 & 0.0960 \\
    B (DML) & 0.0040 & 0.0112 \\
    C (DML+residual) & 0.0050 & 0.0118 \\
    D (three-stage) & 0.0746 & 0.0680 \\
    \bottomrule
  \end{tabular}
\end{table}

\begin{table}[t]
  \centering
  \caption{Normalized RMSEs and tail absolute errors (validation sample; $N=500$, seed=42).}
  \label{tab:scale-metrics}
  \setlength{\tabcolsep}{4pt}
  \renewcommand{\arraystretch}{1.05}
  \resizebox{\linewidth}{!}{%
  \begin{tabular}{lcccccccc}
    \toprule
    Model & nRMSE$_P$ & relRMSE$_P$ & nRMSE$_\Delta$ & nRMSE$_\Gamma$ & $|e_P|_{90}$ & $|e_P|_{99}$ & $|e_\Gamma|_{90}$ & $|e_\Gamma|_{99}$ \\
    \midrule
    A (price-only) & 0.0015 & 0.0027 & 0.0123 & 0.2525 & 0.000379 & 0.000696 & 1.100 & 7.794 \\
    B (DML) & 0.0018 & 0.0034 & 0.0079 & 0.2378 & 0.000459 & 0.000906 & 0.3532 & 3.628 \\
    C (DML+residual) & 0.0018 & 0.0034 & 0.0079 & 0.2378 & 0.000458 & 0.000902 & 0.3504 & 3.628 \\
    D (three-stage) & 0.0018 & 0.0033 & 0.0078 & 0.2376 & 0.000458 & 0.000879 & 0.2503 & 3.628 \\
    \bottomrule
  \end{tabular}
  }
\end{table}
We additionally report scale-invariant metrics and tail error quantiles in Table~\ref{tab:scale-metrics}. For price RMSE, we use two scale-free metrics:
\[
\mathrm{nRMSE}_P:=\frac{\mathrm{RMSE}_P}{\sqrt{\mathbb{E}[(u^{\mathrm{ref}})^2]}},
\qquad
\mathrm{relRMSE}_P:=\frac{\mathrm{RMSE}_P}{\mathbb{E}[u^{\mathrm{ref}}]}.
\]
For Greeks we use
\[
\mathrm{nRMSE}_\Delta:=\frac{\mathrm{RMSE}_\Delta}{\sqrt{\mathbb{E}[(\Delta^{\mathrm{ref}})^2]}},
\qquad
\mathrm{nRMSE}_\Gamma:=\frac{\mathrm{RMSE}_\Gamma}{\sqrt{\mathbb{E}[(\Gamma^{\mathrm{ref}})^2]}}.
\]
The tail metrics $|e_P|_{p}$ and $|e_\Gamma|_{p}$ denote the $p$-th percentile of the absolute price and gamma errors, respectively.
Tables~\ref{tab:variant-global}--\ref{tab:scale-metrics} suggest three points. First, Model~B improves delta accuracy (first-order risk) and vega accuracy relative to the price-only fit (Model~A) (Table~\ref{tab:variant-global}). For gamma $\Gamma$, improvements are clearer in tail error quantiles than in global RMSE (Table~\ref{tab:scale-metrics}). Second, the one-stage PIDE residual penalty (Model~C) does not improve prices or Greeks relative to Model~B. This is consistent with jump models, where the residual can be reduced via cancellation between the differential and jump terms rather than by learning an interpretable jump operator. Third, the three-stage model (Model~D) yields the best overall $\Delta$ and vega RMSE among the Greek-supervised models and slightly improves $\Gamma$ tail error quantiles, although $\Gamma$ remains the most delicate target in the region $(|x|<0.05,\, \tau\le 1/252)$.

\begin{figure}[!htbp]
  \centering
  \includegraphics[width=0.92\textwidth]{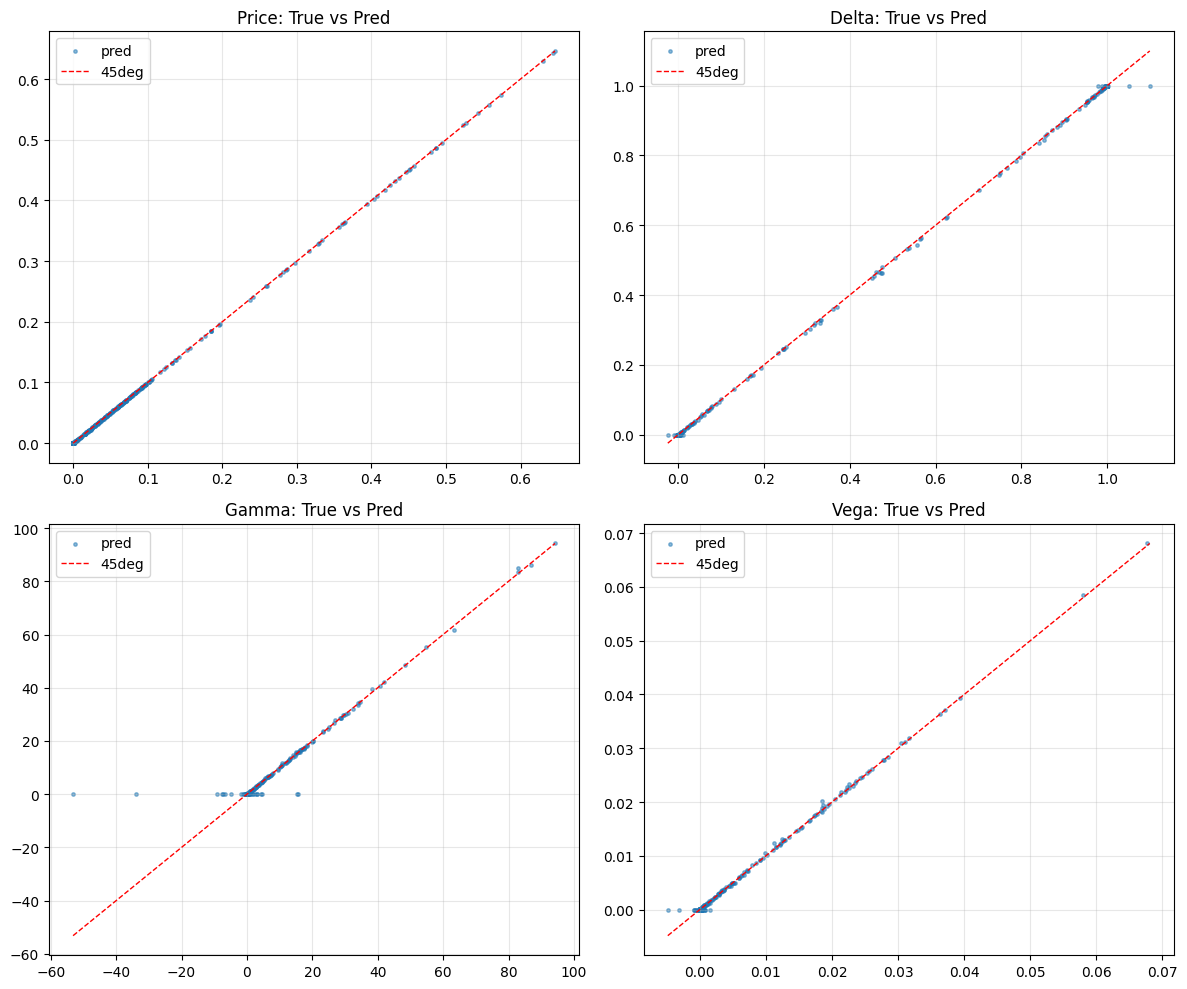}
  \caption{Three-stage model: true vs predicted (price, delta, gamma, vega) on validation.}
  \label{fig:pred-scatter}
\end{figure}

Figure~\ref{fig:pred-scatter} visualizes true vs predicted prices and Greeks for Model~D. In the ultra-short maturity regime ($\tau\to 0$), $\Gamma$ is extremely localized around the money: for the majority of (deep ITM/OTM) evaluation points the true $\Gamma$ is numerically close to zero, so the scatter plots naturally show a dense cluster at $\Gamma\approx 0$.

\subsection{Jump-term comparison}
\begin{figure}[!htbp]
  \centering
  \begin{subfigure}[t]{0.49\textwidth}
    \centering
    \includegraphics[width=\textwidth]{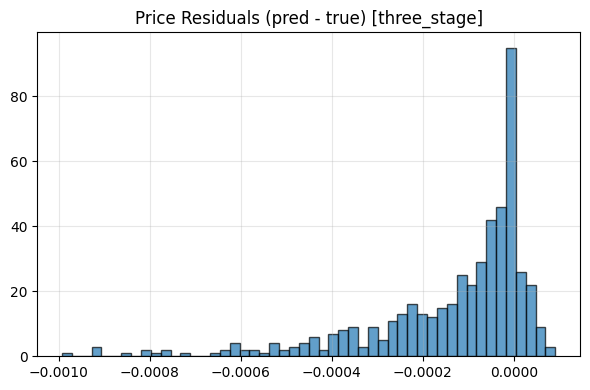}
    \caption{Price residuals $\widehat{u}-u$.}
  \end{subfigure}
  \hfill
  \begin{subfigure}[t]{0.49\textwidth}
    \centering
    \includegraphics[width=\textwidth]{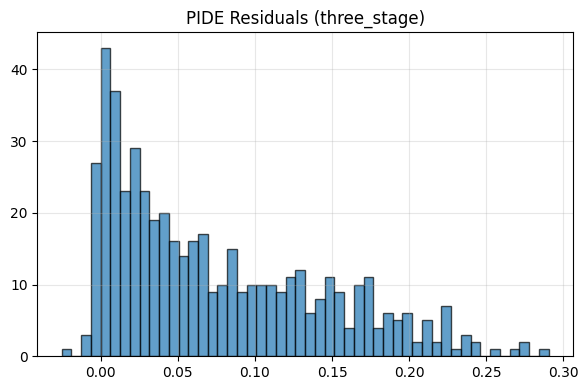}
    \caption{PIDE residual $R(\mathbf{x})$ (three-stage).}
  \end{subfigure}
  \caption{Error distributions for the three-stage model.}
  \label{fig:residuals}
\end{figure}

\begin{figure}[t]
  \centering
  \begin{subfigure}[t]{0.48\linewidth}
    \centering
    \includegraphics[width=\linewidth]{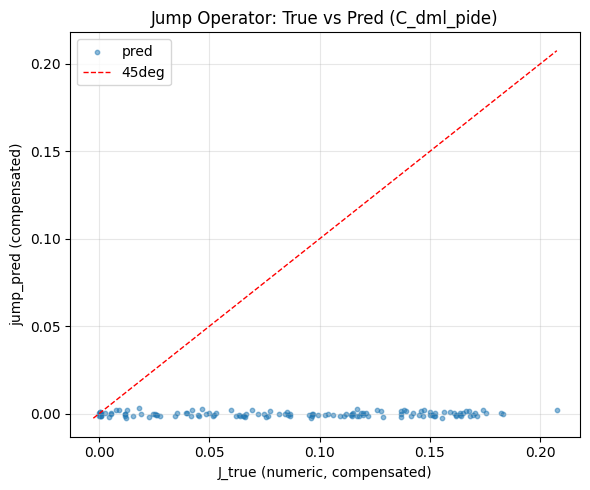}
    \caption{C (DML+residual)}
  \end{subfigure}
  \hfill
  \begin{subfigure}[t]{0.48\linewidth}
    \centering
    \includegraphics[width=\linewidth]{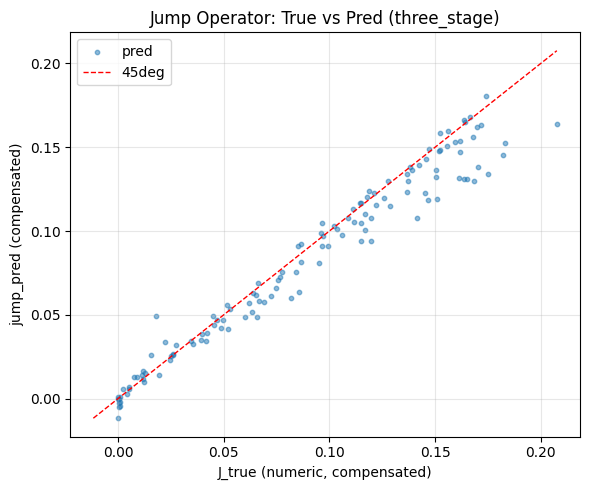}
    \caption{D (three-stage)}
  \end{subfigure}
  \caption{Jump-term check in the data domain ($|x|\le 0.5$): predicted compensated jump term vs numeric integral proxy.}
  \label{fig:jump-sanity}
\end{figure}

This subsection clarifies why a small PIDE residual alone is not evidence that the learned jump component represents the intended compensated jump integral.
We first report error distributions for the final three-stage model (Figure~\ref{fig:residuals}), and then directly test whether the learned compensated jump contribution $J_\psi$ matches a numerical proxy computed from the reference pricer (Figure~\ref{fig:jump-sanity}).

We focus on the domain $|x|\le 0.5$. Despite its small PIDE residual, the residual-regularized model C (DML+residual) without jump supervision shows a large mismatch between $J_\psi$ and $J^{\text{ref}}$ (RMSE $=9.34\times 10^{-2}$).
The three-stage model yields a smaller jump-term error (RMSE $=1.35\times 10^{-2}$).
Price RMSE on the same points is comparable (C: $7.17\times 10^{-3}$, D: $5.89\times 10^{-3}$). Therefore, residual magnitude alone is not sufficient for model selection when diffusion and jump terms can offset each other.

This indicates that when the jump contribution is a free network output, it can absorb approximation errors from the diffusion part and still drive the residual toward zero.
Supervising $J_\psi$ against a numerical proxy (three-stage schedule) improves identification of the jump contribution.

\subsection{One-day delta-hedging}\label{sec:delta-hedging}
\begin{figure}[!htbp]
  \centering
  \includegraphics[width=0.62\textwidth]{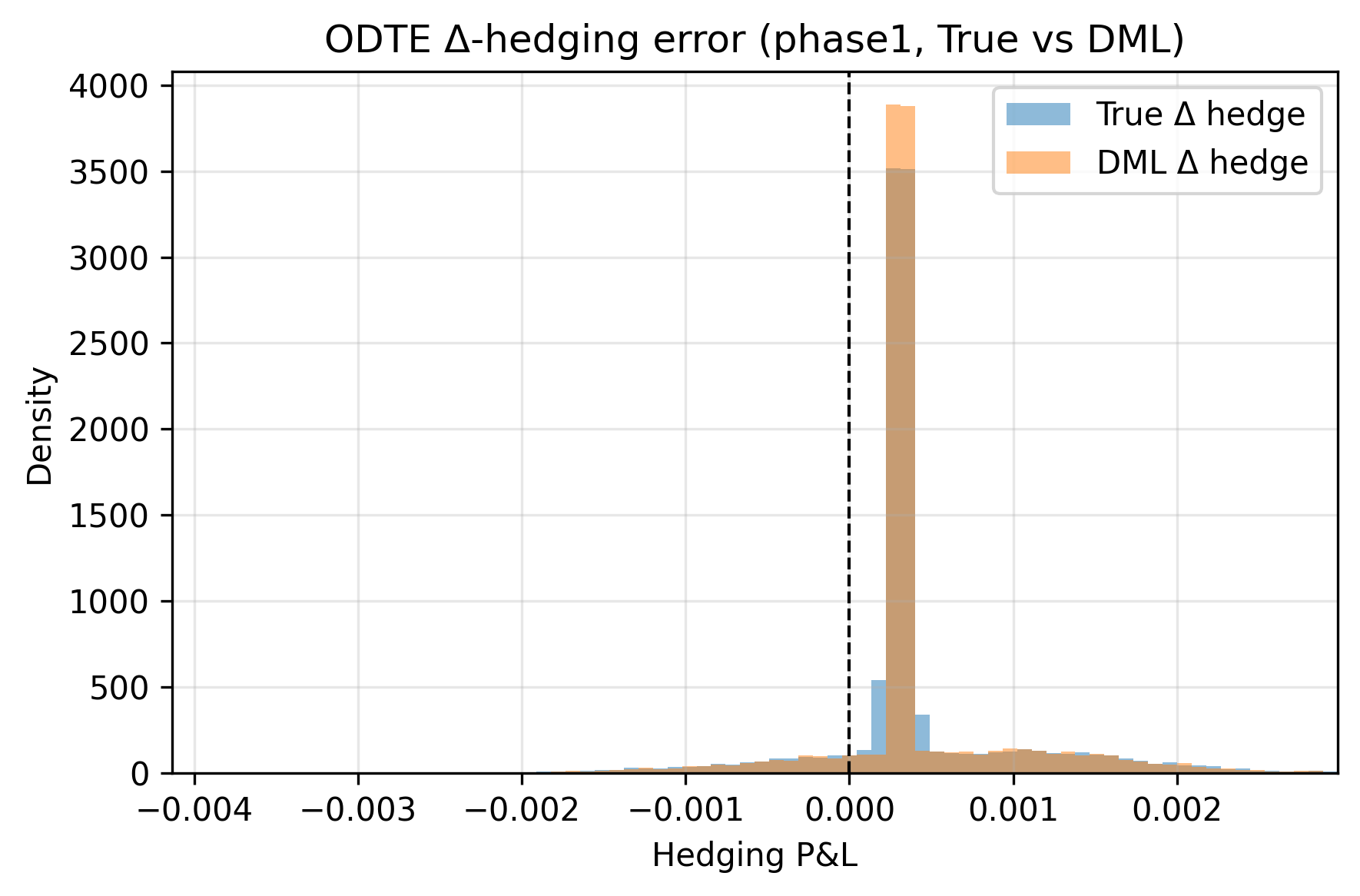}
  \caption{One-day $\Delta$-hedging P\&L (True vs DML delta), stressed Bates parameter regime.}
  \label{fig:hedge-delta}
\end{figure}

The data-generating process is the Bates model, and we simulate $n_{\text{paths}}=5000$ paths of $(S_t,V_t)$ over one trading day $T=1/252$ using an Euler scheme with $n_{\text{steps}}=24$.
Jumps are simulated by a Bernoulli approximation with step probability $p\approx \lambda\,\Delta t$.
To avoid unrealistically large per-step jump probabilities when $\lambda\,\Delta t$ is not very small, we cap this probability at $0.2$, i.e. we use $p=\min(\lambda\,\Delta t,0.2)$.
We use a stressed parameter set
\[
(v_0,\kappa,\theta,\sigma_v,\rho,\lambda,\mu_J,\sigma_J)=(0.04,3.0,0.04,1.0,-0.8,2.0,-0.05,0.20),
\]
chosen to amplify jump activity and gamma effects.
First, we conduct stock-only $\Delta$ hedges, where the hedge portfolio contains the underlying and a cash account. We run the hedge separately for each strike $K\in\{0.9,1.0,1.1\}$ on the same set of simulated $(S_t,V_t)$ paths, yielding 5000 P\&L outcomes per strike. At $t=0$, we short one call with strike $K\in\{0.9,1.0,1.1\}$ and maturity $T$.
At each rebalance time $t_k$, we compute delta from either the Fourier pricer (benchmark) or the DML model, rebalance the stock position, and carry the resulting cash balance at rate $r$. At maturity we record hedging P\&L as the final portfolio value minus the option payoff.
Aggregating across strikes ($15000$ paths total), we report summary statistics in Table~\ref{tab:delta-hedge-summary}. Figure~\ref{fig:hedge-delta} plots P\&L histograms, showing that the distributional difference between true and model delta hedges is small.

\begin{table}[!htbp]
\centering
\caption{One-day $\Delta$-hedging P\&L summary.}
\label{tab:delta-hedge-summary}
\begin{tabular}{lccccc}
\toprule
Model & Mean & Std & 5\% & Median & 95\% \\
\midrule
True $\Delta$ & $-1.40\times 10^{-4}$ & $9.92\times 10^{-3}$ & $-5.60\times 10^{-4}$ & $+3.20\times 10^{-4}$ & $+1.63\times 10^{-3}$ \\
DML $\Delta$  & $-1.30\times 10^{-4}$ & $9.92\times 10^{-3}$ & $-5.00\times 10^{-4}$ & $+3.50\times 10^{-4}$ & $+1.59\times 10^{-3}$ \\
\bottomrule
\end{tabular}
\end{table}

Conditioning on whether at least one price jump occurs along the underlying path, we find much larger dispersion for the DML hedge on jump samples.
Specifically, $\mathrm{sd}(\mathrm{P\&L}_{\text{DML}})\approx 0.02263$ on jump samples ($N=150$ pooled path--strike observations) versus $\approx 0.00971$ on no-jump samples ($N=14{,}850$).
This confirms that realized jumps substantially thicken the one-day hedge-P\&L tails in this stressed experiment.

Second, we compare three one-option second-order hedges:
(i) a ratio-type $\Delta+\Gamma$ hedge;
(ii) a weighted ridge least-squares (LS) hedge that fits $(\Delta,\Gamma)$ jointly across stock and one hedge option; and
(iii) a P\&L-increment regression hedge (PL-LS) that learns state-dependent hedge ratios with ridge regularization.

Here ``ratio-type'' means that the hedge-option position is chosen from an explicit gamma ratio with ridge regularization, and the stock position is then set to neutralize the remaining delta. Such a hedge, using only one additional option, can be ill-conditioned when the hedge option's gamma is small, because the hedge weight effectively scales like a ratio of gammas.

The LS and PL-LS hedges use the same instrument set but choose the positions jointly from a penalized fitting criterion or from a regression on price increments.

Concretely, for a given hedge strike $K_{\mathrm{H}}$, let $C^{\mathrm{main}}_t$ and $C^{\mathrm{H}}_t$ denote the prices of the liability option and the hedge option, and let $(\Delta^{\mathrm{main}}_t,\Gamma^{\mathrm{main}}_t)$ and $(\Delta^{\mathrm{H}}_t,\Gamma^{\mathrm{H}}_t)$ denote their spot Greeks.
For the self-financing portfolio
\[
\Pi_t=-C^{\mathrm{main}}_t+w_{\mathrm{H},t}C^{\mathrm{H}}_t+w_{S,t}S_t+B_t,
\]
the ratio-type $\Delta+\Gamma$ implementation updates the hedge-option position at each rebalance date by the ridge-stabilized gamma match
\[
w_{\mathrm{H},t}=\frac{\Gamma^{\mathrm{main}}_t\,\Gamma^{\mathrm{H}}_t}{(\Gamma^{\mathrm{H}}_t)^2+\eta_{\Gamma}},
\]
which approximates $\Gamma^{\mathrm{main}}_t/\Gamma^{\mathrm{H}}_t$ when $|\Gamma^{\mathrm{H}}_t|$ is not too small.
The stock position is then chosen as
\[
w_{S,t}=\Delta^{\mathrm{main}}_t-w_{\mathrm{H},t}\Delta^{\mathrm{H}}_t,
\]
so that the portfolio is approximately both delta- and gamma-neutral.
At inception we set
\[
B_0=C^{\mathrm{main}}_0-w_{\mathrm{H},0}C^{\mathrm{H}}_0-w_{S,0}S_0,
\]
and at each rebalance date update $(w_{\mathrm{H},t},w_{S,t})$ and the cash account so that the strategy remains self-financing.
The weighted ridge LS hedge uses the same instrument pair $(S_t,C^{\mathrm{H}}_t)$, but chooses $(w_S,w_{\mathrm{H}})$ by minimizing
\[
\min_{w_S,w_{\mathrm{H}}}\;
w_\Delta(\Delta^{\mathrm{main}}_t-w_S-w_{\mathrm{H}}\Delta^{\mathrm{H}}_t)^2
+w_\Gamma(\Gamma^{\mathrm{main}}_t-w_{\mathrm{H}}\Gamma^{\mathrm{H}}_t)^2
+\lambda(w_S^2+w_{\mathrm{H}}^2).
\]

PL-LS estimates time-dependent coefficient vectors $(A_k,B_k)$ from a ridge regression of option price increments $\Delta C_{\text{main}}$ on basis-weighted $(\Delta S,\Delta C_{\text{hedge}})$ and then sets the hedge using $w_S=\phi^\top A_k$ and $w_{\mathrm{H}}=\phi^\top B_k$ (with $\phi=[1,\log m,(\log m)^2,V,\tau]^\top$).

\begin{figure}[!htbp]
\centering
\includegraphics[width=\linewidth]{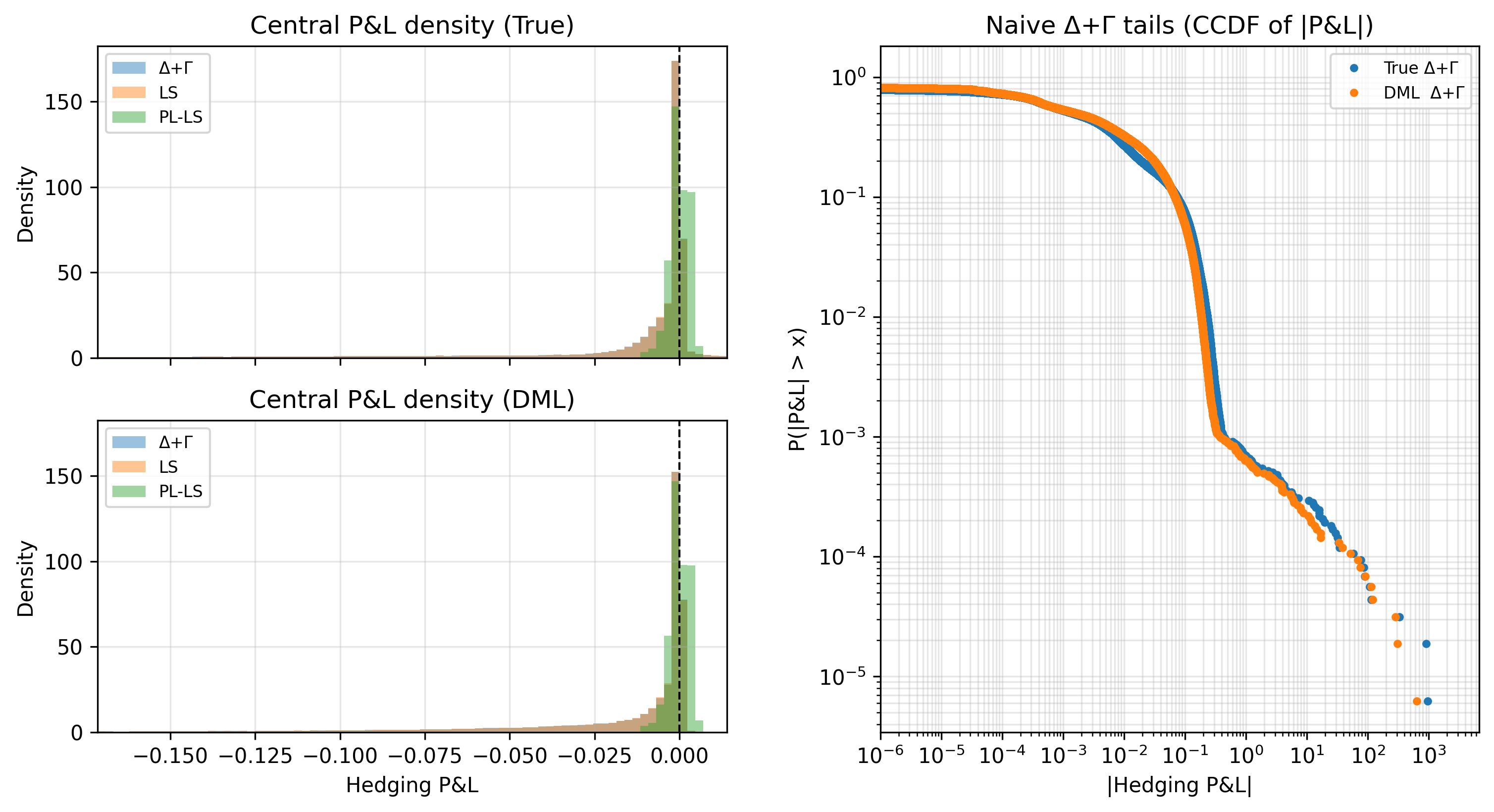}
\caption{$\Delta+\Gamma$ hedging illustration for a one-day ATM liability call ($K_{\mathrm{main}}=1.00$). In each run the hedge portfolio contains the stock and one additional one-day call at a single strike $K_{\mathrm{H}}\in\{0.99,1.00,1.01,1.02\}$; the four $K_{\mathrm{H}}$ choices are run separately and the plotted distributions pool the resulting outcomes across them. Left: central-region density of hedging P\&L for ratio-type $\Delta+\Gamma$, LS, and PL-LS, shown separately for (top) true Greeks and (bottom) DML Greeks. The two central-density panels are almost indistinguishable at this scale, reflecting that the DML Greeks are close to the benchmark. Right: tail behavior of the ratio-type $\Delta+\Gamma$ hedge, shown as the CCDF of $|\mathrm{P\&L}|$ (true vs.\ DML) on log scales; a one-option gamma-based ratio hedge can become unstable when the hedge option's gamma is near zero.}
\label{fig:hedge-dg}
\end{figure}

We fix the liability option as a one-day ATM call with $K_{\mathrm{main}}=1.00$ and $T=1/252$. We then run four separate hedging experiments, each using the stock plus one additional same-maturity call with hedge strike $K_{\mathrm{H}}\in\{0.99,1.00,1.01,1.02\}$. Figure~\ref{fig:hedge-dg} pools the resulting P\&L observations across these four $K_{\mathrm{H}}$ choices.

Figure~\ref{fig:hedge-dg} summarizes the comparison: the left column shows the central-region P\&L density for the ratio-type $\Delta+\Gamma$, LS, and PL-LS hedges, plotted separately for (top) true Greeks and (bottom) DML Greeks.
The two central-density panels are visually similar at this scale, consistent with the close agreement between the DML and benchmark Greeks. In addition, our DML pricer outputs $(u,\Delta,\Gamma)$ in a single forward pass, so these quantities can be computed without re-pricing loops or finite differences.

The right panel shows the tail of the ratio-type $\Delta+\Gamma$ hedge via the CCDF of $|\mathrm{P\&L}|$.
LS and PL-LS are omitted on the right because their tails collapse near zero at this scale.
Appendix~\ref{app:modelchoice} reports additional model-choice checks.
In particular, a Merton jump-diffusion baseline improves substantially over Black--Scholes when prices are generated under the Bates model, while an SVCJ extension with contemporaneous variance jumps is almost indistinguishable from the Bates model.

\subsection{Calibration experiments}\label{sec:synthetic-calibration}
In the spirit of \cite{SridiBilokon2023} and \cite{PolalaHientzsch2023}, we use the trained DML model as a fast pricing surrogate in calibration, but, for simplicity, calibrate only $\beta=(v_0,\theta,\lambda)$ while keeping the remaining parameters fixed. All calibration targets in this subsection are European calls with $S_0=1$, $r=0.01$, $q=0$, and strikes parameterized by log-moneyness $x=\log(S_0/K)$. For a panel $\{(x_i,\tau_i,P_i^{\mathrm{obs}})\}_{i=1}^N$, we solve the weighted least-squares problem with L-BFGS-B
\begin{equation}
\widehat{\beta}
\;=\;
\arg\min_{\beta\in\mathcal{B}}
\sum_{i=1}^N
w_i\,
\Big(
P^{\mathrm{model}}(x_i,\tau_i;\beta,\bar\eta)-P_i^{\mathrm{obs}}
\Big)^2,
\label{eq:calib-objective}
\end{equation}
with $v_0,\theta\in[0.005,0.5]$ and $\lambda\in[0.01,2.0]$. Unless otherwise noted, we use
\[
w_i = 3^{\mathbf 1\{|x_i|<0.1\}}\,3^{\mathbf 1\{\tau_i<1/252\}}
\]
for the calibration weights. In Experiments~2--4, we also assign the short-maturity weight when $\tau_i=1/252$. Thus near-the-money calls receive weight $3$, short-maturity calls receive weight $3$, and calls satisfying both conditions receive weight $9$.

\paragraph{Experiment 1.}
The call panel contains $68$ prices with parameters $(v_0,\theta,\lambda)=(0.04,0.04,0.30)$. The calls lie on a log-moneyness grid $x\in[-0.5,0.5]$ with $17$ equally spaced points and maturities $\tau\in\{0.25,0.50,0.75,1.00\}/252$. Replacing the direct Fourier pricer with the trained DML surrogate makes calibration about $90\times$ faster. The surface fit remains accurate: the DML calibration yields price RMSE $1.53\times 10^{-5}$, versus $1.23\times 10^{-7}$ for direct Fourier calibration. The fitted parameters, however, differ more noticeably. The DML calibration settles at $\hat\lambda=0.20$, while direct Fourier calibration recovers $\hat\lambda\approx 0.30$. Even direct Fourier calibration does not recover the generating vector exactly, returning $\hat\theta=0.05$. On such a short-maturity panel, exact parameter recovery is therefore difficult even when pricing error is tiny, a point reinforced by the identification results in Experiment~3.

\begin{figure}[H]
  \centering
  \includegraphics[width=0.82\linewidth]{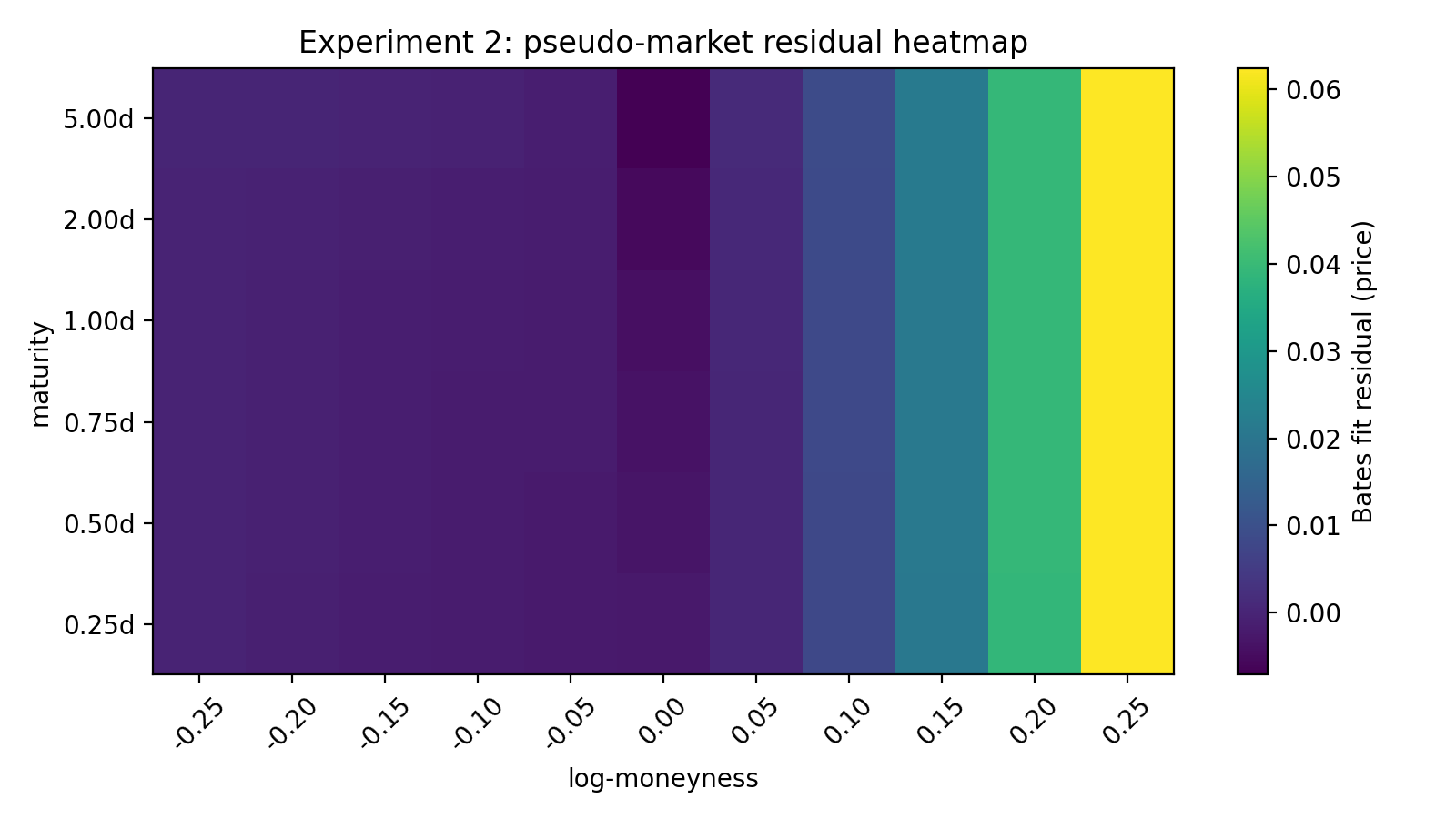}
  \caption{Residual heatmap from Fourier pricer calibration to a smooth pseudo-market surface. We use $66$ Black--Scholes calls on the grid $x\in[-0.25,0.25]$ with $11$ equally spaced points and $\tau\in\{0.25,0.50,0.75,1.00,2.00,5.00\}/252$.}
  \label{fig:calib-exp2}
\end{figure}

\paragraph{Experiment 2.}
We next examine how model misspecification manifests itself in calibration. The implied-volatility surface is specified as
\[
\sigma_{\mathrm{pm}}(x,\tau)=a(\tau)\bigl(1-b(\tau)x+c(\tau)x^2\bigr),
\]
with $a(\tau)=0.18+0.03e^{-35\tau}+0.005\sin(150\tau)$, $b(\tau)=0.20+0.10e^{-15\tau}$, and $c(\tau)=0.08+0.05e^{-20\tau}$. We convert this surface to call prices, floor volatilities at $5\%$ and cap them at $100\%$ pointwise, and then fit the Bates model by Fourier-pricer calibration under the box constraints in \eqref{eq:calib-objective}. Figure~\ref{fig:calib-exp2} shows structured residuals across moneyness and maturity. The fit pushes $\hat\theta$ to its lower bound $0.005$, $\hat\lambda$ to its lower bound $0.01$, and $\hat v_0$ to $0.0073$. This pattern indicates that the optimizer suppresses both variance and jump activity in order to mimic a surface that lies outside the model class.
 
\begin{figure}[H]
  \centering
  \includegraphics[width=0.78\linewidth]{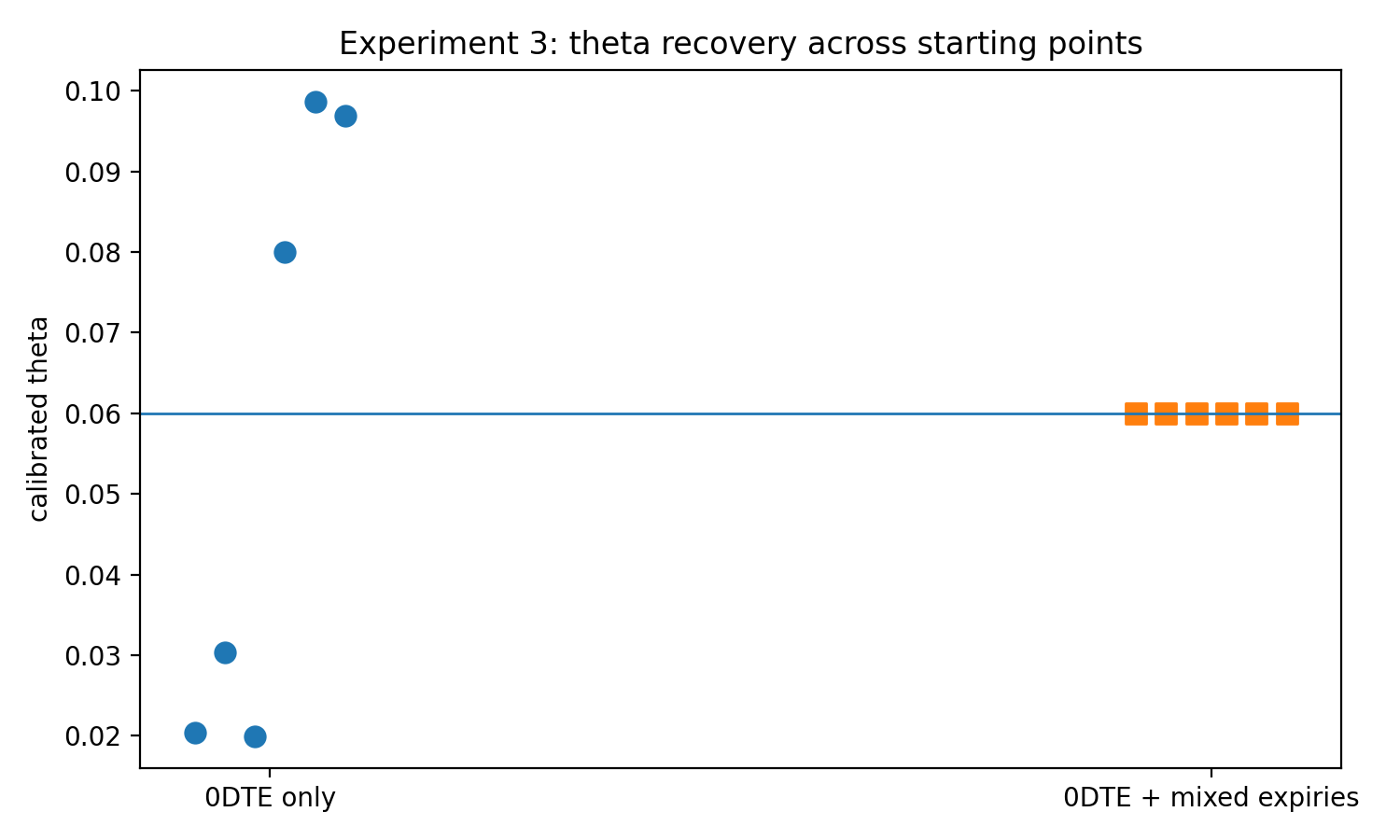}
  \caption{Calibrated $\theta$ across six starting values. We generate calls with parameters $(v_0,\theta,\lambda)=(0.04,0.06,0.30)$ on the common grid $x\in[-0.25,0.25]$ with $11$ equally spaced points. We compare a $44$ 0DTE-only panel with $\tau\in\{0.25,0.50,0.75,1.00\}/252$ against an $88$ mixed-expiry panel that additionally includes $\tau\in\{2.00,5.00,10.00,20.00\}/252$.}
  \label{fig:calib-exp3}
\end{figure}

\paragraph{Experiment 3.}
Here we show how longer expiries improve identification of slower-moving stochastic-volatility parameters. Starting from $(0.03,0.03,0.15)$, the 0DTE-only calibration recovers $(\hat v_0,\hat\theta,\hat\lambda)=(0.0401,0.0304,0.3021)$. By contrast, the mixed-expiry panel recovers $(0.0400,0.0600,0.3000)$. Figure~\ref{fig:calib-exp3} reports the corresponding multistart results across six starting values, including $(0.03,0.03,0.15)$. The 0DTE-only fits produce $\hat\theta$ values in the range $[0.019,0.099]$, whereas the mixed-expiry fits concentrate in $[0.059,0.060]$.

\paragraph{Experiment 4.}
Inspired by the parameter-sensitivity supervision in \cite{ZhangAmiciMorandotti2025}, we consider a parameter-gradient extension focused on jump intensity. We train two DML networks on calls generated by the Bates model: one on prices only, and one on prices together with the price sensitivity with respect to $\lambda$.
Letting $\xi_j$ denote the full network input for training sample $j$, we train the sensitivity-augmented network with
\begin{equation}
\mathcal L_\phi^{(\lambda)}
=
\frac{1}{M}\sum_{j=1}^M
\big(
\widehat P_\phi(\xi_j)-P_j
\big)^2
+
\omega_\lambda
\frac{1}{M}\sum_{j=1}^M
\Big(
\partial_\lambda \widehat P_\phi(\xi_j)-g_j^{(\lambda)}
\Big)^2,
\label{eq:lambda-training-loss}
\end{equation}
where $\omega_\lambda=20$ and $g_j^{(\lambda)}$ is the jump-intensity price-sensitivity label generated by automatic differentiation. Calibration still solves weighted least squares in price space:
\begin{equation}
\widehat{\lambda}
=
\arg\min_{\lambda\in[0.01,2.0]}
\sum_{i=1}^N
w_i
\Big(
\widehat P_\phi(x_i,\tau_i;\lambda)-P_i^{\mathrm{obs}}
\Big)^2.
\label{eq:lambda-calib-objective}
\end{equation}
Table~\ref{tab:lambda-pilot} shows that Fourier pricer calibration recovers $\hat\lambda=0.59997$ for $\lambda_{\mathrm{true}}=0.60$. The price-only DML proxy overestimates jump intensity, yielding $\hat\lambda=0.73878$, whereas the price+$\lambda$-sensitivity DML moves the estimate to $\hat\lambda=0.56889$. We use the same mixed-expiry grid as Calibration Experiment 3 so that the jump-gradient calibration is not evaluated on a distinct expiry panel. The sensitivity-supervised surrogate does not exactly recover $\lambda$, but it materially reduces the absolute error from about $0.139$ to about $0.031$. For ultra-short maturities, $\lambda$ is the parameter most directly tied to jump exposure, so supervising $\partial_\lambda P$ appears helpful for calibrating that direction, even though it does not solve the full multi-parameter calibration problem by itself.

\begin{table}[H]
  \centering
  \caption{Gradient-based calibration in the jump-intensity direction. The revised replication script uses the same mixed-expiry grid as Calibration Experiment~3, $\tau\in\{0.25,0.50,0.75,1.00,2.00,5.00,10.00,20.00\}/252$, with $x\in[-0.5,0.5]$ on $21$ points.}
  \label{tab:lambda-pilot}
  \begin{tabular}{lcc}
    \toprule
    Method & $\hat\lambda$ & $|\hat\lambda-\lambda_{\mathrm{true}}|$ \\
    \midrule
    Fourier pricer calibration & 0.59997 & $3.33\times 10^{-5}$ \\
    Price-only DML & 0.73878 & $1.39\times 10^{-1}$ \\
    Price+$\lambda$-sensitivity DML & 0.56889 & $3.11\times 10^{-2}$ \\
    \bottomrule
  \end{tabular}
\end{table}

\subsection{Extension to rough volatility}\label{sec:rough-extension}
Here we ask whether the Bates/SVJD-B DML architecture can be extended to rough volatility. We replace the Markovian state $(s,v,\tau)$ with an $M$-factor Markovian approximation of rough volatility, so that the pricing network uses $(s,y_1,\ldots,y_M,\tau)$ as input \cite{AbiJaberElEuch2019}. Specifically, we approximate the fractional kernel
\begin{equation}
K_H(t)=\frac{t^{H-1/2}}{\Gamma(H+1/2)}
\end{equation}
by a finite sum of exponentials,
\begin{equation}
K_H(t) \approx K_M(t)=\sum_{m=1}^M c_m e^{-\gamma_m t}
\end{equation}
where the coefficients $c_m$ are estimated by nonnegative least squares. Let
\[
  u=u(s,y_1,\ldots,y_M,\tau),\qquad V^M(y)=v_0+\sum_{m=1}^M y_m,
\]
and define
\[
  b_m(y)=-\gamma_m y_m+c_m\kappa(\theta-V^M(y)).
\]
The $M$-factor rough diffusion generator is
\begin{align}\label{eq:rough-finitefactor-generator}
\mathcal A_M^{\mathrm{rough}} u
=&\,(r-q)s u_s+
\sum_{m=1}^M b_m(y)u_{y_m}
+
\frac12 V^M s^2u_{ss} \notag\\
&+\rho\xi V^M s\sum_{m=1}^M c_m u_{s y_m}
+
\frac12\xi^2V^M
\sum_{m=1}^M\sum_{\ell=1}^M c_m c_\ell u_{y_my_\ell}.
\end{align}
The jump-network target is the compensated price-jump operator per unit intensity,
\begin{equation}\label{eq:rough-jump-label}
\mathcal J^{\mathrm{label}}(s,y,\tau)
=
\mathbb E_Z\{u(se^Z,y,\tau)-u(s,y,\tau)\}
-\kappa_J s u_s(s,y,\tau),
\qquad
Z\sim N(\mu_J,\sigma_J^2),
\end{equation}
where $\kappa_J=\mathbb E[e^Z-1]$. The reference prices are computed using the rough-volatility variance-reduction method of McCrickerd and Pakkanen~\cite{McCrickerdPakkanen2018}, combined with Merton's lognormal-jump mixture~\cite{Merton1976}. As in the Bates/SVJD-B experiment, Gauss--Hermite quadrature is used only to construct supervised labels for this compensated jump operator under the reference model.

The pricing-equation residual uses the learned jump block $\widehat{\mathcal J}_\phi$:
\begin{equation}\label{eq:rough-finitefactor-residual}
  \mathcal R^M_{\theta,\phi}
  =\partial_\tau \widehat u_\theta
  -\mathcal A_M^{\mathrm{rough}}\widehat u_\theta
  -\lambda\widehat{\mathcal J}_\phi
  +r\widehat u_\theta 
\end{equation}
and the price network is trained on inputs $(\log(S/K),y_1,\ldots,y_M,\tau)$ with a loss function similar to \eqref{eq:loss}:
\begin{equation}\label{eq:rough-dml-loss}
\mathcal L_{\mathrm{RHM}}
=\mathcal L_{\mathrm{price}}
+\omega_\Delta \mathcal L_\Delta
+\omega_\Gamma \mathcal L_\Gamma
+\omega_J\mathcal L_{\mathrm{jump}}
+\omega_R\,\mathbb E\{(\mathcal R^M_{\theta,\phi})^2\}
+\omega_{\mathrm{NA}}\,\mathbb E\!\left[\mathcal P_{\mathrm{NA}}^M(x,y,\tau)\right].
\end{equation}
The term $\mathcal L_{\mathrm{jump}}$ trains the learned jump block $\widehat{\mathcal J}_\phi$ against \eqref{eq:rough-jump-label} and $\mathcal P_{\mathrm{NA}}^M$ denotes the rough-state analogue of the no-arbitrage penalty $\mathcal P_{\mathrm{NA}}$ in \eqref{eq:loss}, applied to the $M$-factor RHM-DML price surface $\widehat u_\theta(x,y,\tau)$.

We generate a volatility surface from a higher-dimensional RHM with $(M,H)=(12,0.07)$ and fit Bates/SVJD-B, SVCJ, and a lower-dimensional RHM approximation with $(M,H)=(4,0.10)$. Table~\ref{tab:rough-mismatch} reports implied-volatility RMSEs under three calibration grids. The RHM-with-jumps specification attains the lowest implied-volatility RMSE in all three grids, supporting the interpretation that rough memory helps reproduce the shape of the implied-volatility surface when combined with jumps.

\begin{table}[H]
  \centering
    \caption{Implied-volatility errors. Calibration uses $\log(S/K)\in\{-0.18,-0.09,0,0.09,0.18\}$ and the maturity grid indicated in the first column: 0DTE-only $=\{0.25,0.50,0.75,1.00\}/252$, 0DTE+5D $=\{0.25,0.50,0.75,1.00,2.00,5.00\}/252$, and mixed-expiry $=\{0.25,0.50,0.75,1.00,2.00,5.00,10.00,20.00\}/252$. The evaluation grid used to compute the errors is $\log(S/K)\in\{-0.14,-0.04,0.04,0.14\}$ and $\tau\in\{1.50,5.00\}/252$.}
  \label{tab:rough-mismatch}
  \footnotesize
  \setlength{\tabcolsep}{7pt}
  \begin{tabular}{@{}llc@{}}
    \toprule
    Calibration grid & Candidate & IV RMSE \\
    \midrule
    0DTE-only & Bates/SVJD-B & $9.841\times10^{-2}$ \\
               & SVCJ         & $9.841\times10^{-2}$ \\
               & RHM with jumps & $4.834\times10^{-2}$ \\
    \addlinespace[2pt]
    0DTE+5D   & Bates/SVJD-B & $2.630\times10^{-1}$ \\
               & SVCJ         & $2.630\times10^{-1}$ \\
               & RHM with jumps & $3.777\times10^{-2}$ \\
    \addlinespace[2pt]
    Mixed-expiry & Bates/SVJD-B & $2.474\times10^{-1}$ \\
                 & SVCJ         & $2.476\times10^{-1}$ \\
                 & RHM with jumps & $3.558\times10^{-2}$ \\
    \bottomrule
  \end{tabular}
\end{table}

Table~\ref{tab:finitefactor-rhm-dml} reports the accuracy of the finite-factor RHM-DML model under the residual-regularized DML framework. Price, delta, and jump-operator errors remain small, whereas gamma has the largest tail error because it is a second derivative concentrated around the 0DTE near-ATM region.

\begin{table}[H]
  \centering
  \caption{$M$-factor RHM-DML accuracy ($M=4$). Accuracy is measured against a conditional-Merton Monte Carlo reference model. The jump operator is the compensated per-unit-intensity jump contribution learned by $\widehat{\mathcal J}_\phi$.}
  \label{tab:finitefactor-rhm-dml}
  \setlength{\tabcolsep}{5pt}
  \begin{tabular}{lcccc}
    \toprule
    Quantity & RMSE & Mean abs. error & q95 abs error & q99 abs error \\
    \midrule
    Price & $6.80\times10^{-4}$ & $5.69\times10^{-4}$ & $1.31\times10^{-3}$ & $1.52\times10^{-3}$ \\
    Delta & $7.42\times10^{-3}$ & $5.32\times10^{-3}$ & $1.40\times10^{-2}$ & $2.53\times10^{-2}$ \\
    Gamma & $4.61\times10^{-1}$ & $2.15\times10^{-1}$ & $1.24$ & $1.85$ \\
    Jump operator & $7.78\times10^{-4}$ & $6.23\times10^{-4}$ & $1.67\times10^{-3}$ & $2.22\times10^{-3}$ \\
    \bottomrule
  \end{tabular}
\end{table}

Table~\ref{tab:rough-mismatch-hedge} reports results for terminal one-step delta hedges. We choose the initial delta at $t=0$ and simulate $10{,}000$ terminal payoffs for a one-day ATM call with $\tau=1/252$. The Black--Scholes IV delta yields the smallest mean absolute P\&L and RMSE, but the remaining tail losses indicate that a single stock-delta hedge cannot remove jump risk. Perfect hedging in rough Heston models is theoretically possible when the forward variance curve is available as a hedging instrument~\cite{ElEuchRosenbaum2018}; more applied work formulates rough-volatility hedging through additional instruments or finite-factor partial hedging~\cite{FukasawaHorvathTankov2021,MotteHainaut2024}. Survey evidence also emphasizes that rough-volatility models may be valuable for pricing and risk management, but that practical implementation remains numerically and methodologically demanding~\cite{HirakiShinozaki2024}. Rough models without jumps can struggle to reproduce wide-moneyness implied-volatility smiles, motivating a rough-plus-jump rather than a pure-rough comparison~\cite{BandiFusariGazzaniReno2026}.

\begin{table}[H]
  \centering
  \caption{Terminal stock-delta hedging results. Mean $|P\&L|$ is the sample mean of absolute terminal hedge P\&L; q95 and q99 denote the 95th and 99th percentiles of absolute terminal hedge P\&L. The fitted deltas are evaluated using the model parameters calibrated from the 0DTE+5D row of Table~\ref{tab:rough-mismatch}. Bates/SVJD-B and SVCJ have identical rounded values. The $M$-factor RHM-DML delta reproduces the fitted RHM-MC delta.}
  \label{tab:rough-mismatch-hedge}
  \setlength{\tabcolsep}{4pt}
  \begin{tabular}{lccccc}
    \toprule
    Hedge input & Delta & Mean $|P\&L|$ & RMSE & q95 abs & q99 abs \\
    \midrule
    Unhedged & $0.000$ & $5.07\times10^{-3}$ & $7.94\times10^{-3}$ & $1.14\times10^{-2}$ & $1.89\times10^{-2}$ \\
    Black--Scholes IV delta & $0.504$ & $3.44\times10^{-3}$ & $5.92\times10^{-3}$ & $9.21\times10^{-3}$ & $2.16\times10^{-2}$ \\
    Bates/SVJD-B or SVCJ fitted delta & $0.519$ & $3.48\times10^{-3}$ & $6.01\times10^{-3}$ & $9.49\times10^{-3}$ & $2.24\times10^{-2}$ \\
    $M$-factor RHM-DML delta & $0.578$ & $3.69\times10^{-3}$ & $6.44\times10^{-3}$ & $1.07\times10^{-2}$ & $2.56\times10^{-2}$ \\
    \bottomrule
  \end{tabular}
\end{table}

\section{Conclusion}
We develop a differential machine learning framework for 0DTE options under the Bates model. The method embeds the price in a Black--Scholes representation with a maturity-gated variance correction and combines joint price-and-Greek supervision with jump-aware PIDE regularization.
Across the numerical experiments, the three-stage design improves identification of the jump term while keeping price accuracy competitive, delta and vega errors low, and inference substantially faster than the Fourier benchmark. The calibration exercises show both the value and the limits of the approach: the network is a useful fast surrogate inside weighted least-squares calibration, but short-dated panels can still suffer from misspecification and weak parameter identification. Supervising jump-intensity sensitivity further improves recovery in this direction, reducing the jump-intensity calibration error relative to a price-only surrogate. The framework can also be extended to rough volatility models through a multi-factor Markovian approximation. Future research will extend the framework from simulation to market data.

\section*{Declaration of AI-assisted technologies}
During the preparation of this manuscript, the author used generative AI and AI-assisted tools to assist with manuscript drafting, language refinement, code drafting and debugging, and discussion of the presentation of the research. The author directed the project, reviewed, revised, and verified the manuscript and code, and takes full responsibility for the final manuscript.

\bibliographystyle{apalike}

\appendix

\section{Model comparison}\label{app:modelchoice}
Because 0DTE options are extremely short-dated, it is natural to ask whether simpler dynamics would suffice for the pricing and hedging checks considered in this paper.
This appendix reports two model-comparison checks:
(i) Black--Scholes and Merton jump-diffusion baselines versus the Bates model, and
(ii) an SVCJ-type extension with contemporaneous jumps in price and variance, specified within the affine jump-diffusion framework of \cite{DuffiePanSingleton2000}.

\subsection{Black--Scholes and Merton versus the Bates model}

We use the Fourier pricer as a benchmark and compare two analytic baselines:
(i) Black--Scholes with constant volatility $\sigma=\sqrt{v_0}$ and no jumps, and
(ii) Merton's lognormal jump-diffusion \cite{Merton1976} with the \emph{same} jump parameters $(\lambda,\mu_J,\sigma_J)$ and constant diffusion volatility $\sqrt{v_0}$.
The parameter set is the stressed regime used in the hedging experiment in the main text:
\[
(v_0,\kappa,\theta,\sigma_v,\rho,\lambda,\mu_J,\sigma_J)
=(0.04,3.0,0.04,1.0,-0.8,2.0,-0.05,0.20),
\]
with $r=1\%$ and $q=0$.

\begin{table}[t]
  \centering
  \caption{Pricing error of Black--Scholes and Merton relative to Bates benchmark prices, evaluated on a 164-point $(x,\tau)$ grid. The second column reports RMSE on the core short-dated bucket $(|x|<0.05,\ \tau\le 1/252)$ (92 grid points, including repeated endpoints).}
  \label{tab:bs-merton-vs-bates}
  \begin{tabular}{lcc}
    \toprule
    Model & Price RMSE (global) & Price RMSE $(|x|<0.05,\ \tau\le 1/252)$ \\
    \midrule
    Black--Scholes & 0.000303 & 0.000297 \\
    Merton & 0.000182 & 0.000121 \\
    \bottomrule
  \end{tabular}
\end{table}

Merton's jump-diffusion substantially reduces pricing error relative to Black--Scholes (Table~\ref{tab:bs-merton-vs-bates}), indicating that even overnight options load on jump risk.

We also run a one-day discrete-time $\Delta$ hedge under simulated Bates dynamics (using the same setup as Section~\ref{sec:delta-hedging}) and compute hedge deltas using BS, Merton, and the Bates model.
Table~\ref{tab:hedge-bs-merton} reports summary statistics, both unconditionally and conditional on whether at least one jump occurs along the path.
Jump realizations dominate tail outcomes and cannot be hedged away by delta hedging; differences across delta models are therefore small relative to jump-driven tail risk in this setup.

\begin{table}[t]
  \centering
  \caption{One-day discrete-time $\Delta$-hedging P\&L under simulated Bates dynamics. ``jump'' means the path contains at least one price jump during the day; ``no-jump'' means no price jump occurs. The last column reports a nonparametric 95\% bootstrap CI for CVaR$_{1\%}$.}
  \label{tab:hedge-bs-merton}
  \resizebox{\linewidth}{!}{%
  \begin{tabular}{llrrrrl}
    \toprule
    Model & Subset & $n$ & mean & sd & q01 & CVaR01 [95\% CI] \\
    \midrule
    BS & all & 30{,}000 & 0.000084 & 0.007247 & -0.001641 & -0.0339 [-0.042,-0.026] \\
    BS & jump & 237 & -0.040952 & 0.070066 & -0.293528 & -0.3683 [-0.456,-0.251] \\
    BS & no-jump & 29{,}763 & 0.000411 & 0.000568 & -0.001274 & -0.0020 [-0.0021,-0.0019] \\
    Bates & all & 30{,}000 & 0.000075 & 0.007254 & -0.001728 & -0.0340 [-0.043,-0.026] \\
    Bates & jump & 237 & -0.041152 & 0.069999 & -0.287896 & -0.3677 [-0.456,-0.248] \\
    Bates & no-jump & 29{,}763 & 0.000403 & 0.000616 & -0.001473 & -0.0021 [-0.0021,-0.0020] \\
    Merton & all & 30{,}000 & 0.000083 & 0.007238 & -0.001651 & -0.0339 [-0.042,-0.026] \\
    Merton & jump & 237 & -0.040979 & 0.069925 & -0.292678 & -0.3680 [-0.456,-0.250] \\
    Merton & no-jump & 29{,}763 & 0.000410 & 0.000574 & -0.001304 & -0.0020 [-0.0021,-0.0019] \\
    \bottomrule
  \end{tabular}
  }
\end{table}

\subsection{SVCJ vs Bates: do volatility jumps matter for 0DTE?}
To assess whether volatility jumps matter for 0DTE options, we consider an SVCJ-type extension (stochastic volatility with contemporaneous jumps in price and variance) in which price and variance jump simultaneously at Poisson times.
Within the affine jump-diffusion framework of \cite{DuffiePanSingleton2000}, we specify the log-price jump as $Y\sim\mathcal{N}(\mu_J,\sigma_J^2)$ and the variance jump as $Z\sim \mathrm{Exp}(\text{mean}=\mu_{vJ})$, with $Z$ independent of $Y$ but arriving with the same intensity $\lambda$.
We set $\mu_{vJ}=0.02$ (a positive variance-jump mean), and keep $(v_0,\kappa,\theta,\sigma_v,\rho,\lambda,\mu_J,\sigma_J)$ equal to the stressed regime above.
Table~\ref{tab:svcj-vs-bates-price} shows that the incremental pricing effect of adding variance jumps is small. For very short maturities, vega exposure is limited; variance jumps can therefore be weakly identified unless longer expiries are included in calibration/training.
The evaluation grid here uses four same-day maturity slices, $\tau\in\{0.25,0.50,0.75,1.00\}\times(1/252)$.
Accordingly, the metric ``ATM'' aggregates all grid points with $|x|<0.1$ across all four slices, whereas ``ATM+short'' additionally imposes $\tau<1/252$, i.e.\ it retains only the first three slices.

\begin{table}[t]
  \centering
  \caption{Incremental pricing effect of adding variance jumps, reported as the price difference $C^{\mathrm{SVCJ}}-C^{\mathrm{Bates}}$ and summarized by its RMSE over several evaluation subsets ($\mu_{vJ}=0.02$). All metrics are evaluated on the same 164-point $(x,\tau)$ grid.}
  \label{tab:svcj-vs-bates-price}
  \begin{tabular}{lcccc}
    \toprule
    Metric & global & ATM & short & ATM+short \\
    \midrule
    Price RMSE & $3.16\times 10^{-7}$ & $3.47\times 10^{-7}$ & $1.93\times 10^{-7}$ & $2.11\times 10^{-7}$ \\
    \bottomrule
  \end{tabular}
\end{table}

We conduct one-day delta hedging under simulated SVCJ dynamics and compare hedges based on Bates vs SVCJ deltas. Table~\ref{tab:hedge-svcj} shows that the P\&L distributions are very similar at the reported precision.

\begin{table}[t]
  \centering
  \caption{One-day $\Delta$-hedging P\&L under simulated SVCJ dynamics: summary statistics comparing deltas from Bates vs SVCJ (negative values indicate profit). The last column reports a nonparametric 95\% bootstrap CI for CVaR$_{1\%}$.}
  \label{tab:hedge-svcj}
  \resizebox{\linewidth}{!}{%
  \begin{tabular}{llrrrrl}
    \toprule
    Model & Subset & $n$ & mean & sd & q01 & CVaR01 [95\% CI] \\
    \midrule
    Bates & all & 9{,}000 & -0.000316 & 0.012868 & -0.002961 & -0.0734 [-0.103,-0.054] \\
    Bates & jump & 108 & -0.058557 & 0.101288 & -0.358808 & -0.4407 [-0.521,-0.301] \\
    Bates & no-jump & 8{,}892 & 0.000392 & 0.001140 & -0.002165 & -0.0028 [-0.0029,-0.0026] \\
    SVCJ & all & 9{,}000 & -0.000316 & 0.012868 & -0.002961 & -0.0734 [-0.101,-0.047] \\
    SVCJ & jump & 108 & -0.058557 & 0.101288 & -0.358808 & -0.4407 [-0.521,-0.301] \\
    SVCJ & no-jump & 8{,}892 & 0.000392 & 0.001140 & -0.002165 & -0.0028 [-0.0029,-0.0026] \\
    \bottomrule
  \end{tabular}
  }
\end{table}

\end{document}